\documentstyle[12pt]{article}
\vspace{1 cm}
\def\lapproxeq{\lower .7ex\hbox{$\;\stackrel{\textstyle <}{\sim}\;$}}
\def\gapproxeq{\lower .7ex\hbox{$\;\stackrel{\textstyle >}{\sim}\;$}}

\begin{document}

\titlepage

\begin{flushright} RAL-94-028 \\ March 1994
\end{flushright}

\begin{center}
\vspace*{2cm}
{\large{\bf Direct photoproduction of jets in the $k_T$-factorisation
prescription}}
\end{center}

\vspace*{.75cm}
\begin{center}
J.R.\ Forshaw and R.G.\ Roberts \\
Rutherford Appleton Laboratory, \\ Chilton, Didcot OX11 0QX, England. \\
\end{center}

\vspace*{1.5cm}

\begin{abstract}
We study the cross section for dijet production via direct photons, as seen at
the HERA collider. Rather than assuming the standard factorisation formula for
the cross section (where the incoming partons are assumed to be on-shell), we
use the $k_T$-factorisation formula. This formalism ought to be more
appropriate for jet production in regions where the cross section is
sensitive to the small-$x$ parton distribution functions. We study the rapidity
distribution and the azimuthal angular distribution of the jets and compare
with the traditional approach. Our calculations reveal a breakdown of the
traditional approach as one moves into the regime of small $x$.
\end{abstract}

\newpage
In a previous paper \cite{p1}, we illustrated the importance of studying the
cross section for dijet production via direct photons at the HERA collider in
order to gain information regarding the gluon content of the proton at
small-$x$. Since that paper was published, the ZEUS collaboration has shown
that it is possible to separate their events into `direct photon' events and
`resolved photon' events using a cut on the fraction of the photon energy
which enters the hard scatter \cite{ZEUS}. This encourages us to study further
this important process and so, in this paper, we wish to return to the
theoretical evaluation of the dijet cross section in the context of the
`$k_T$-factorisation' prescription, which ought to be used for small
$x \sim p_T^2/s$ ($p_T$ is a typical jet transverse momentum and $s$ is the
$e p$ centre-of-mass energy).

At large $x \sim p_T^2/s$, we know that the $\ln p_T$ terms, which are present
in the perturbative expansion of the cross section, can become significant and
need to be summed -- leading to the violation of the scaling predictions of the
parton model. The summation of the leading logarithms (which arise as a result
of multi-parton branching before the hard scatter) is performed using the DGLAP
(Dokshitzer, Gribov, Lipatov, Altarelli, Parisi) equations \cite{DGLAP}. The
infra-red singularities (reflecting the presence of long distance physics)
which arise in the calculation of these QCD corrections can be separated from
the short distance contributions and thus we are led to cross sections which
factorise (schematically) in the following way \cite{fact}:
\begin{equation}
\sigma \sim F(...,\mu^2) \otimes \hat{\sigma}(...,\mu^2,Q^2)
\label{qfact}
\end{equation}
The function $F$ contains the (unknown) long-distance physics whilst only
short distance physics is contained in the `hard' subprocess cross section
$\hat{\sigma}$. It is usual to choose the factorisation scale $(\mu^2)$, to
equal the hard scale $(Q^2)$, in order to subsume the leading logarithms
within the function $F$. Central to this factorisation formula is the
assumption that the `hard' cross section can be computed assuming that the
incoming partons have negligible transverse momenta (and virtualities)
compared with the scale of the hard process. We shall call this the
`$Q^2$-factorisation' approach.

For small enough $x$ we expect that the $\ln x$ terms, which are present
in the perturbative expansion, will become dominant. We refer
to the summation of these leading logarithms in $1/x$ as the BFKL
(Balitsky, Fadin, Kuraev, Lipatov) summation \cite{BFKL}.
If the logarithms in $x$ are indeed dominant, then we are no longer
allowed to use the formula of eqn.(\ref{qfact}). We must use the
$k_T$-factorisation formula of refs.\cite{Catani,CE,Russians}, which states
(again schematically) that
\begin{equation}
\sigma \sim F(...,k_{Ti}^2,\mu^2) \otimes \hat{\sigma}(...,k_{Ti}^2,\mu^2,Q^2)
\label{ktfact}
\end{equation}
The scales, $k_{Ti}$, are the transverse momenta carried by the partons
entering the hard scatter.

In the specific case of direct photoproduction of dijets the
$Q^2$-factorisation formula can be written:
\begin{equation}
\frac{d^3\sigma^{dir}}{dy_1 dy_2 dp_T^2} \; = \;
zf_{\gamma /e}(z)\; xg(x,Q^2)\; \frac{d \hat \sigma(\gamma g \rightarrow
q \bar{q})} {d \hat t}  \label{qxsecn}
\end{equation}
The final state parton rapidities are denoted $y_1$ and $y_2$, and $p_T$
denotes their transverse momentum in the $\gamma p$ centre-of-mass frame.
The azimuthal angular distribution is trivial since the jets are produced
back-to-back. The photon flux is determined by the function,
$f_{\gamma/e}(z)$. The long distance physics is contained in the gluon
structure function, $xg(x,Q^2)$, and the short distance physics is contained
in the cross section:
\begin{equation}
\frac{d \hat \sigma(\gamma g \rightarrow q \bar{q})} {d \hat t} =
\frac{2 \pi \alpha \alpha_s}{\hat{s}^2} \left( \frac{\hat{t}}{\hat{u}} +
\frac{\hat{u}}{\hat{t}} \right)
\end{equation}
For simplicity we show only the gluonic contribution to the cross section (it
is this contribution which dominates for jets produced in the electron
direction, i.e. small $x$). The subprocess Mandelstam variables are denoted
$\hat{s}, \hat{t}$ and $\hat{u}$ and satisfy the usual kinematic relations.

The corresponding formula in the language of $k_T$-factorisation can be written
thus \cite{Russians}:
\begin{equation}
\frac{d^2\sigma^{dir}}{dy_1dy_2} \; = \; \int \frac{d^2{\mathrm p_T}}{\pi} \,
\frac{d^2{\mathrm k_T}}{\pi} \, d^2{\mathrm p_T'} \,
\delta({\mathrm p_T} + {\mathrm p_T'} - {\mathrm k_T}) \,
{\cal J}(\nu,p_T^2,k_T^2,\theta) \, zf_{\gamma /e}(z)\; \Phi(x,k_T^2) \;
\frac{d \hat \sigma(\gamma g^{*} \rightarrow q \bar{q})} {d \hat t}
\end{equation}
The final state partons have transverse momentum vectors ${\mathrm p_T}$ and
${\mathrm p_T'}$ which are no longer
equal and opposite for non-vanishing gluon momentum,
${\mathrm k_T}$. We can use the delta function to perform the ${\mathrm p_T'}$
integrals. One of the two remaining angular integrals can also be performed
(since the integrand only depends upon $\theta$, the angle between
${\mathrm k_T}$ and ${\mathrm p_T}$). We can therefore write:
\begin{equation}
\frac{d^2\sigma^{dir}}{dy_1dy_2} \; = \; \int dp_T^2 \, dk_T^2 \,
\frac{d\theta}{2 \pi} \, {\cal J}(\nu,p_T^2,k_T^2,\theta) \,
zf_{\gamma /e}(z)\; \Phi(x,k_T^2)\; \frac{d \hat \sigma(\gamma g^{*}
\rightarrow q \bar{q})} {d \hat t}  \label{ktxsecn}
\end{equation}
For the `hard scatter' cross section we use \cite{HM}:
\begin{equation}
\frac{d \hat \sigma(\gamma g^* \rightarrow q \bar{q})} {d \hat t} =
\frac{2 \pi \alpha \alpha_s}{(\hat{s}+k_T^2)^2} \left( \frac{\hat{t}}{\hat{u}}
+ \frac{\hat{u}}{\hat{t}} - \frac{2k_T^2 \hat{s}}{\hat{u}\hat{t}} + \frac{12
k_T^2 \hat{s}}{(\hat{u} + \hat{t})^2} \right)
\end{equation}
${\cal J}$ is the Jacobean and $\nu = \exp\,(y_1 - y_2)$.
Some important kinematic relations follow.
\begin{equation}
p_T'^2 = p_T^2 + k_T^2 - 2 k_T \,p_T \, \cos\,\theta
\end{equation}
The momentum fractions $x$ and $z$ can then be written
\begin{eqnarray}
x &=& \frac{1}{2 E_p} (p_T e^{-y_1} + p_T' e^{-y_2}) \nonumber \\
z &=& \frac{1}{2 E_e} (p_T e^{ y_1} + p_T' e^{ y_2})
\end{eqnarray}
and
\begin{equation}
-\hat{t} = p_T^2 + p_T p_T' / \nu
\end{equation}
Also, $\hat{s} = 4 x z E_e E_p$ and $\hat{s}+\hat{t}+\hat{u} + k_T^2 = 0$.
We can write the Jacobean explicitly as:
\begin{equation}
xz {\cal{J}}(\nu,p_T^2,k_T^2,\theta) = \frac{p_T p_T'}{4\nu E_e E_p} \left[
\nu^2 + \nu \left( \frac{p_T'}{p_T} + \frac{\Delta^2}{p_T p_T'} \right) +
\frac{\Delta^2}{p_T^2}\right]
\end{equation}
where $\Delta^2 = (p_T^2 + p_T'^2 - k_T^2)/2$.

To regain eq.(\ref{qxsecn}) from eq.(\ref{ktxsecn}), one assumes that the
integrand can be approximated by its value at $k_T^2 = 0$ (the $\theta$
integral is then trivial) and that the maximum value of $k_T^2 = Q^2$. This
leads to the relation:
\begin{equation}
xg(x,Q^2) = \int_0^{Q^2} dk_T^2 \, \Phi(x,k_T^2)
\end{equation}
We can re-arrange eq.(\ref{ktxsecn}) in order to examine the azimuthal
distribution of the final state partons:
\begin{equation}
\frac{d^3\sigma^{dir}}{dy_1 dy_2 d\phi/2\pi} \; = \; \int dp_T^2 \, dk_T^2 \,
 {\cal J}(\nu,p_T^2,k_T^2,\theta) \, \frac{p_T' \cos\,\phi}{k_T \cos\,
\theta - p_T \sin^2\phi} zf_{\gamma /e}(z)\; \Phi(x,k_T^2)\;
\frac{d \hat \sigma(\gamma g^{*} \rightarrow q \bar{q})} {d \hat t}
\end{equation}
The angle between ${\mathrm p_T}$ and ${\mathrm p_T'}$ is $\phi$ and
\begin{equation}
\cos\,\theta = \frac{p_T}{k_T}\sin^2\phi \pm \left( 1 -
\frac{p_T^2}{k_T^2}\sin^2\phi \right)^{1/2} \cos\,\phi
\end{equation}
In the limit of $k_T^2 \rightarrow 0$, the integrand becomes a delta function
at the only allowed point in phase space, i.e. where $\phi = \pi$.

At this point we should emphasise that if one intends to use the standard DGLAP
evolution equations to determine the $Q^2$-evolution of $xg$ then it is
formally inconsistent to construct $\Phi$ by taking the derivative and then
using the $k_T$-factorisation formula above. In the DGLAP formalism, the QCD
corrections to the cross section contain mass
singularities, which must be removed by factorisation. This should be
contrasted with the BFKL formalism where there are no mass singularities.
However, we acknowledge that the absence of such singularities does not
guarantee sensible results and if our cross section depends critically on
inherently infra-red physics we may not be able to trust the results.

These issues are dealt with nicely in the paper of Collins and Ellis \cite{CE}.
They introduce a factorisation scale to isolate the infra-red physics and,
working in moment space, i.e.
\begin{equation}
\tilde{\Phi}(j,k_T^2;\mu^2) = \int_0^1 dx \, x^{j-1} \Phi(x,k_T^2;\mu^2)
\end{equation}
they find that
\begin{equation}
\tilde{\Phi}(j,k_T^2;\mu^2) = \gamma_c(j)\; \frac{1}{k_T^2} \left(
\frac{k_T^2}{\mu^2} \right)^{\gamma_c(j)} x\tilde g(j,\mu^2)
\end{equation}
where $\gamma_c(j)$ is the BFKL anomalous dimension function which takes on the
value of (exactly) $1/2$ when $j = 1 + 12 \alpha_s \ln 2  / \pi$. The BFKL
corrections which multiply the gluon distribution are calculated by solving
the BFKL equation with an infra-red cut-off of $\mu$.  To obtain the final
cross section, $\Phi$ must be convoluted with the `impact
factor' -- in our case this leads to the expression:
\begin{eqnarray}
\frac{d^2\sigma^{dir}}{dy_1dy_2} \; &=& \; \int dp_T^2 \, zf_{\gamma/e}(z)
\, xg(x,\mu^2) \, \frac{d \hat \sigma(\gamma g \rightarrow q \bar{q})}
{d \hat t} \nonumber \\ &+& \int dp_T^2 \, \int_{\mu^2} dk_T^2 \,
\frac{d\theta}{2 \pi} \, {\cal J}(\nu,p_T^2,k_T^2,\theta) \,
zf_{\gamma /e}(z)\; \Phi(x,k_T^2;\mu^2)\; \frac{d \hat \sigma(\gamma g^{*}
\rightarrow q \bar{q})} {d \hat t}
\end{eqnarray}

In this letter, we do not wish to focus on the specific solution for the
function, $\Phi$. Rather, we want to investigate the general features of this
new factorisation prescription -- and to ask whether it might be expected to
produce significant deviations from the more standard formalism.

To this end, let us use the following BFKL motivated form for the gluon
distribution function (we ignore quarks throughout this paper -- they
are not included in the BFKL formalism since they do not lead to the dominant
small $x$ behaviour).
\begin{equation}
\Phi(x,k^2) = {\cal N} \frac{x^{-\lambda}}{k^2}(k^2)^{1/2} \exp \left(
-\frac{\ln(k^2/k_0^2)}{2\lambda'' \ln (x_0/x + 1)} \right)
\end{equation}
where
\begin{eqnarray}
\lambda  &=& \frac{3 \alpha_s}{\pi} 4 \ln 2 \nonumber \\
\lambda''&=& \frac{3 \alpha_s}{\pi} 28 \zeta(3)
\end{eqnarray}
The parameters, $x_0$ and $k_0^2$, are treated as essentially
free -- and we show results for different choices. Let us briefly reflect upon
our prejudices regarding their likely values.
The scale, $k_0^2$, is determined by the
size of the hadron, and we therefore expect rather small values, $\sim
\Lambda_{QCD}^2$. In the limit of $x \rightarrow 0$, eq.(18) reduces to
that which is predicted by the BFKL formalism. However, in order to ensure
sensible behaviour at intermediate values of $x$, we modify the asymptotic
logarithm, i.e. $\ln 1/x \rightarrow \ln(x_0/x + 1)$. This modification ensures
a narrow distribution in $k^2$ at large $x$ (consistent with the
$Q^2$-factorisation assumption that the partons are on-shell). One can think
of $x_0$ as the parameter which delineates the onset of BFKL behaviour from
the more traditional large $x$ behaviour, which suggests $x_0 \sim 0.01$.
The normalisation, ${\cal N}$, is a constant and unimportant for the purposes
of this paper.

The following figures show the effect of $k_T$-factorisation on the cross
section for direct photoproduction of dijets at HERA, i.e.
820 GeV protons colliding with 30 GeV electrons. The cross sections are
$ep$ cross sections, and the photon flux was approximated using the same
Weiszacker-Williams formula as in ref.\cite{p1}. We took $\lambda = 0.5$
and $\lambda'' = 6.1$, in eq.(19) but allowed a running coupling (with
$\Lambda_{QCD} = 200$ MeV and $Q^2 = p_T^2/4$) in the hard scattering cross
section.

In figs.(1), we show the variation of the cross section with the jet rapidity
(the jets are fixed to have equal rapidity, i.e. $y_1 = y_2 = y$) for four
different combinations of the parameters $k_0$ and $x_0$. In all cases, we
find that the $k_T$-factorisation prescription (eq.(17)) produces a cross
section which is significantly smaller than that which is obtained using the
$Q^2$-factorisation prescription (eq.(3)) for those jets which are produced in
the electron direction, i.e. small $x$. Not surprisingly, the effect increases
as $k_0$ and $x_0$ are increased -- since we move the peak of the
$k^2$-distribution to larger energies as $k_0$ increases and broaden the
width of the $k^2$-distribution as $x_0$ increases.

The transition from the $k_T$- to $Q^2$-factorisation prescriptions is
illustrated in figs.(2). As the factorisation scale, $\mu^2$, is increased the
cross section tends toward the standard, $Q^2$-factorisation, prescription
result (shown as the horizontal dotted line on each plot). Exact agreement is
never quite obtained since the upper limit on the $k_T$-factorisation integral
is calculated using the exact kinematics (rather than assuming it to be equal
to $Q^2 = p_T^2/4$, which is the hard scale taken in the $Q^2$-factorisation
calculation). In fig.(2a), we show the $\mu^2$ dependence at $y = \pm 1$ using
the BFKL motivated form for $\Phi$ and typical values of $k_0$ and $x_0$. In
fig.(2b), we show the equivalent plots using a $\Phi$ which has been computed
by taking the derivative of the GRV structure function parametrisations for
the proton \cite{GRV}. As we remarked earlier, such an approach is strictly
inconsistent with the BFKL formalism -- one extracts a parton
$k_T$-distribution from the $Q^2$-factorisation approach
(in which the GRV structure functions are derived) --
it is not clear that such an extraction is meaningful. Even so, our results
again show significant deviation from the standard approach, especially in the
backward direction.

{}From the plots of figs.(1) and (2) it should be clear that the essential BFKL
prediction that the incoming partons have non-negligible transverse momenta has
important consequences at HERA. It destroys the notion of universal small $x$
parton distributions in the standard ($Q^2$-factorisation) formalism.
Consequently, it is essential to measure the small-$x$ parton distribution
functions in as many independent processes as is possible, e.g. $xg(x,Q^2)$
from dijets \cite{p1}, $\partial F_2 / \partial \ln Q^2$ \cite{dF2}, $F_L$
\cite{FL} and $J/\Psi$ production \cite{JPsi} -- we {\em do not} expect
universal agreement between all processes.

Another consequence of the non-zero parton $k_T$ is illustrated in fig.(3). The
azimuthal distribution of the final state partons is plotted. Instead of
back-to-back `jets', we expect a de-correlating effect due to the imbalance of
the incoming $k_T$. This could show up experimentally, as a broadening of the
azimuthal distribution for jets produced in the backward direction compared to
those produced in more forward directions. We note that the apparent spikes at
the peaks of the distributions are, of course, not physical and are simply due
to numerical limitations.

In summary, we have performed a quantitative study of the `$k_T$-factorisation'
prescription when applied to a process that is currently being measured at
HERA. Comparison with the traditional `$Q^2$-factorisation' approach shows
that, in the region where small-$x$ gluons within the proton are being probed,
the discrepancy can be sizeable, i.e. $\simeq 20\%$. This deviation arises
essentially due to the relaxation of the assumption that the partons which
enter the hard scatter are on-shell and has the important consequence that we
expect $xg(x,Q^2)$ (for small $x$) to be process dependent. Another signal
for such corrections would be a deviation from the ideal `back-to-back'
configuration in the azimuthal angle, as shown in fig.(3).

\vskip 1.0cm
\noindent{\large\bf Figure Captions}

\begin{itemize}
\item[Fig.\ 1] A comparison of the $Q^2$- and $k_T$-factorisation calculations
of the dijet rapidity distributions. We have chosen $y_1 = y_2 = y$ and define
negative rapidity to be in the proton direction. The normalisation of $\Phi$
was adjusted to reproduce approximately the value of $xg(x,Q^2)$ of GRV
at $x$=0.01 and $Q^2$ = 10 GeV$^2$.

\item[Fig.\ 2] To illustrate the transition from $k_T$- to $Q^2$-factorisation
as the delineating scale, $\mu^2$, is increased (strictly speaking we use
min($\mu^2,Q^2$) as the delineating scale). In each case the prediction
of eq.(3) is described by the dotted line, the first (second) term in
eq.(17) by the long (short) dashed line and their sum by the dot-dashed
line.

\item[Fig.\ 3] The azimuthal distribution of the final state partons,
determined at $y_1 = y_2 = 1.$ and for two different values of $k_0$.
\end{itemize}

\end{document}